\documentclass[prb, twocolumn, aps, amsmath,amssymb,amsfonts]{revtex4}
\usepackage{graphicx}
\usepackage{graphics}
\usepackage{dcolumn}
\usepackage{bm}
\usepackage{amssymb}
\usepackage{amsmath}
\usepackage{amsfonts}
\usepackage{epsfig}

\newcommand {\ket} [1] {| #1 \rangle}
\newcommand {\bkt} [1] {\langle #1 \rangle}

\newcommand {\pd} [2] {\frac{\partial #1}{\partial #2}}
\newcommand {\td} [2] {\frac{d #1}{d #2}}

 \newcommand {\beq}{\begin{equation}}
\newcommand {\eeq}{\end{equation}}
\newcommand {\bea}{\begin{eqnarray}}
\newcommand {\eea}{\end{eqnarray}}
\begin{document}
\title{Anomalous Hall response of topological insulators}
\author{Dimitrie Culcer} 
\affiliation{ICQD, Hefei National Laboratory for Physical Sciences at the Microscale, University of Science and Technology of China, Hefei 230026, Anhui, China}
\author{S. Das Sarma}
\affiliation{Condensed Matter Theory Center, Department of Physics, University of Maryland, College Park, Maryland 20742-4111, USA}
\begin{abstract}
The anomalous Hall effect due to the surface conduction band of 3D topological insulators with an out-of-plane magnetization is \textit{always} dominated by an intrinsic topological term of the order of the conductivity quantum. We determine the contributions due to the band structure, skew scattering, side jump and magnetic impurities on the same footing, demonstrating that the topological term, renormalized due to disorder, overwhelms all other terms, providing an unmistakable signature of $Z_2$ topological order. Uncharacteristically, skew scattering contributes in the Born approximation as well as in the third order in the scattering potential, while in addition to the side-jump scattering term we identify a novel intrinsic side-jump term of a similar magnitude. These, however, never overwhelm the topological contribution.
\end{abstract}
\date{\today}
\maketitle

\section{Introduction}

Topological insulators (TIs) have been an intense focus of research and witnessed impressive experimental breakthroughs in the past years \citep{KaneMele_QSHE_PRL05, Hasan_TI_RMP10, Qi_TI_RMP_10, Qi_TFT_PRB08}. Several materials were predicted to exhibit TI behavior \cite{Zhang_TI_BandStr_NP09}, and experiment has observed TI behavior in Bi Sb \cite{Hsieh_BiSb_QSHI_Nature08}, Bi$_2$Se$_3$,\cite{Urazhdin_PRB04, Xia_Bi2Se3_LargeGap_NP09, Chen_Bi2Se3_GateControlEFWAL_PRL10} Bi$_2$Te$_3$,\cite{Urazhdin_PRB04, Chen_Bi2Te3_Science09, Hsieh_Bi2Te3_Sb2Te3_PRL09} and Sb$_2$Te$_3$ \cite{Hsieh_Bi2Te3_Sb2Te3_PRL09}, with Heusler alloys also recently predicted to belong to this class \cite{Xiao_Heusler_PRL10}. This extraordinary novel state of matter is remarkable for being insulating in the bulk while conducting along the surface, as well as for the absence of backscattering, which leads to expected high mobilities, Klein tunneling, and the suppression of Anderson localization. 

The existence of chiral surface states is well established experimentally through angle-resolved photoemission (ARPES) and scanning-tunneling microscopy (STM) studies, yet in transport their contribution is overwhelmed by the bulk, which, due to unintentional defect-induced doping, is a low-density metal. A key aim of current physics research is the identification of a definitive signature of surface transport. One avenue is doping with magnetic impurities, whose moments align beyond a certain doping concentration and make TIs magnetic, opening a gap \cite{Yu_TI_QuantAHE_Science10}. Recent theories of magnetic TIs include the Kerr and Faraday effects in a thin TI film coupled to a ferromagnet \cite{Tse_TI_GiantMOKE_PRL10} and in a strong external magnetic field \cite{Tse_TI_QuantB_MOKE_PRB10}, spin transfer \cite{Nomura_TI_ElCrgMgnTxt_PRB10}, and anomalous magnetoresistance \cite{Nagaosa_AMR_TopSfc_PRB10}, and edge contributions \cite{Chu_TI_HalfHall_10}, while experiment has observed magnetization dynamics \cite{Wray_TI_MgnDyn_10} and attempted to control surface magnetism \cite{Zhu_TI_SurfMag_10}. The interplay of magnetism, strong spin-orbit coupling, disorder scattering and driving electric fields leads to the anomalous Hall effect (AHE) in magnetic TIs. The AHE has been a mainstay of condensed matter physics, studied extensively over the years \cite{Luttinger_AHE_PR58, Smit_SS_58, Berger_SJ_PRB1970, NozLew_SSSJ_JP73, Crepieux_AHE_KuboDirac_PRB01, Dugaev_AHE_FM_Localization_PRB01, Kovalev_Multiband_AHE_PRB09, Sinitsyn_AHE_KuboStreda_PRB07, Sinitsyn_AHE_Review_JPCM08, Nagaosa_AHE_PRB08, Yang_SJ_SctUniversality_PRB10}, and the scattering mechanisms involved, skew scattering \cite{Smit_SS_58} and side jump \cite{Berger_SJ_PRB1970, NozLew_SSSJ_JP73, Rashba_SJ_FTS08}, are well known to be non-trivial. Furthermore, a fascinating instance of this effect has been predicted to occur in 2D magnetic TIs, where edge states are believed to give rise to a quantized AHE when the chemical potential lies in the magnetization-induced gap between the surface valence and conduction bands. \cite{Yu_TI_QuantAHE_Science10} (See Ref.\ \onlinecite{Tkachov_TI_Transition_PRB11} for a description of the transition between the ordinary and TI regimes in magnetotransport.)

This work is devoted to the experimentally relevant problem of the AHE on the surface of a \textit{three-dimensional} magnetic TI, focusing on the contribution from the surface conduction band. This already challenging problem is complicated by the necessity of a quantum mechanical starting point and of a matrix formulation describing strong band structure spin-orbit coupling, spin-dependent scattering including relativistic corrections, Klein tunneling and Zitterbewegung on the same footing. The fascinating and counterintuitive conclusion of our study is that a topological band structure contribution, known to exist in magnetic spin-orbit coupled systems \cite{Culcer_AHE_PRB03, ZangNagaosa_TI_Monopole_PRB10}, yields the dominant term, overwhelming \textit{all} scattering contributions. The result, which is distinct from the topological magnetoelectric effect \citep{Hasan_TI_RMP10, Qi_TI_RMP_10}, is of the order of the conductivity quantum, independent of the magnetization, and unambiguously reflects $Z_2$ topological order in transport.

The outline of this paper is as follows. In Section II the Hamiltonian of the system is introduced. In Section III the kinetic equation for the density matrix is derived, containing all spin-dependent scattering terms in the first and second Born approximations. This includes the usual scalar scattering term in the first Born approximation, the skew scattering and side-jump scattering terms in the first Born approximation, and the usual skew scattering term in the second Born approximation (of third order in the scattering potential). In Section IV the various nonequilibrium corrections to the density matrix are determined, and their contributions to the AHE are calculated. Comparison is made with existing experiments as well as predictions for future experiments. Section V contains a summary and conclusions.

\section{Hamiltonian}

We study a topological insulator doped with magnetic impurities, which give a net magnetic moment and spin-dependent scattering. The latter are described by $
{H}_\mathrm{mag} ({\bm r}) = {\bm \sigma} \cdot \sum_I \mathcal{V}({\bm r} - {\bm R}_I)\, {\bm s}_I$, where the sum runs over the positions ${\bm R}_I$ of the magnetic ions with ${\bm s}_I$, and the vector of Pauli spin matrices ${\bm \sigma}$ represents the spins of the conduction electrons. We approximate $\mathcal{V}({\bm r} - {\bm R}_I) = J \, \delta({\bm r} - {\bm R}_I)$, with $J$ the exchange constant between the localized moments and the itinerant carriers, and $V$ the sample volume. The magnetic ions are assumed spin-polarized so that ${\bm s}_I = s \, \hat{\bm z}$. Fourier transforming to the crystal momentum representation $\{ \ket{{\bm k}} \}$, with ${\bm k}$ the wave vector, the ${\bm k}$-diagonal term, $H_\mathrm{mag}^{{\bm k} = {\bm k}'} = n_\mathrm{mag}\, J \, s \sigma_z \equiv M \, \sigma_z$, gives the magnetization, with $n_\mathrm{mag}$ the density of magnetic ions. The ${\bm k}$-off-diagonal term leads to spin-dependent scattering
\begin{equation}\label{Hmagod}
\begin{array}{rl}
\displaystyle {H}_\mathrm{mag}^{{\bm k} \ne {\bm k}'} = & \displaystyle \frac{Js}{V} \, \sigma_z \sum_I e^{i({\bm k} - {\bm k}')\cdot{\bm R}_I}.
\end{array}
\end{equation}
We consider zero temperature, so that the splitting due to ${\bm M}$ is resolved, and the Fermi energy $\varepsilon_F \gg |{\bm M}|$ lies in the \textit{surface} conduction band. We require $\varepsilon_F \tau/\hbar \gg 1$ for our theory to be applicable, with $\tau$ the momentum relaxation time. The effective band Hamiltonian is
\begin{equation}
H_{0{\bm k}} = - Ak \, {\bm \sigma} \cdot \hat{\bm \theta} + {\bm \sigma}\cdot{\bm M} \equiv \frac{\hbar}{2} \, {\bm \sigma} \cdot {\bm \Omega}_{\bm k}, 
\end{equation}
where $\hat{\bm \theta}$ is the tangential unit vector in 2D ${\bm k}$-space, and the constant A = 4.1eV$\AA$ for Bi$_2$Se$_3$ \cite{Zhang_TI_BandStr_NP09}. Scalar terms are negligible at relevant transport densities $\approx 10^{13}$cm$^{-2}$  \cite{Zhang_TI_BandStr_NP09, Culcer_TI_Kineq_PRB10}. The eigenenergies are $\varepsilon_\pm = \pm \sqrt{A^2k^2 + M^2}$. We will frequently use $a_k = 2Ak/\hbar\Omega_k$ and $b_k = 2M/\hbar\Omega_k$, so that $a_k^2 + b_k^2 = 1$, and $a_F$ and $b_F$ at $k = k_F$, with $b_F \ll 1$ at usual transport densities. Interaction with a static, uniform electric field ${\bm E}$ is contained in $H_{E, {\bm k}{\bm k}'} = H_{E, {\bm k}{\bm k}'}^{sc} + H_{E, {\bm k}{\bm k}'}^{sj}$, the scalar part arising from the ordinary position operator $H_{E, {\bm k}{\bm k}'}^{sc} = (e{\bm E}\cdot\hat{\bm r})_{{\bm k}{\bm k}'} \openone$, with $\openone$ the identity matrix in spin space, and the side-jump part $H_{E, {\bm k}{\bm k}'}^{sj} = e\lambda \, {\bm \sigma}\cdot({\bm k}\times{\bm E}) \, \delta_{{\bm k}{\bm k}'}$ arising from the spin-orbit modification to the position operator, $\hat{\bm r} \rightarrow \hat{\bm r} + \lambda {\bm \sigma} \times \hat{\bm k}$, \cite{Sinitsyn_AHE_Review_JPCM08} with $\lambda$ a material-specific constant. Elastic scattering off charged impurities, static defects and magnetic ions (but not phonons or electrons) is contained in the scattering potential $U_{{\bm k}{\bm k}'} = \bar{U}_{{\bm k}{\bm k}'} \sum_{J} e^{i({\bm k} - {\bm k}')\cdot{\bm R}_J}$, with ${\bm R}_J$ the random locations of the impurities, which incorporates the magnetic scattering term ${H}_\mathrm{mag}^{{\bm k} \ne {\bm k}'}$ of Eq. (\ref{Hmagod}). The potential of a single magnetic impurity $\bar{U}_{{\bm k}{\bm k}'} = (1 - i \, \lambda \, {\bm \sigma} \cdot {\bm k} \times {\bm k}')\, \mathcal{U}_{{\bm k}{\bm k}'} - (Js/V) \, \sigma_z $, with $\mathcal{U}_{{\bm k}{\bm k}'}$ the matrix element between plane waves, and $\lambda k_F^2 \ll 1$, while $J = 0$ for non-magnetic impurities. The full Hamiltonian $H_{{\bm k}} = H_{0{\bm k}} + H_{E{\bm k}{\bm k}'} + U_{{\bm k}{\bm k}'} + {H}_\mathrm{mag}^{{\bm k} = {\bm k}'}$. We do not include weak localization, which in the regime $\varepsilon_F\tau/\hbar \gg 1$ produce only a small correction to the conductivity, and is known to be unimportant in the AHE. \cite{Dugaev_AHE_FM_Localization_PRB01}

It is well known that in magnetic TIs the Berry curvatures of the surface valence and conduction bands are different, so that when the chemical potential is in the band gap the conductivity is $e^2/2h$. \cite{ZangNagaosa_TI_Monopole_PRB10, Nomura_3DTI_QAHE_PRL11} This quantity is regarded as a constant offset in our work, which is concerned rather with the response of the conduction electrons, that is, the surface conduction band. 

Nontrivial processes such as spin-flip scattering and the Kondo effect can be accounted for in our theory, provided that one works in the grand canonical ensemble in order to account fully for the spin-flip factors. Several reasons compel us to avoid doing so. Given that the magnetization must be large enough to be resolved, the local magnetic moments are assumed to be all lined up. Moreover, although it is reasonable to assume that the Kondo effect may affect the diagonal term in the conductivity significantly, there is no evidence that it should influence the AHE. We will find in this work that all corrections to the AHE are zeroth order or higher in the scattering potential, in other words next-to-leading order. Considering the very few studies of the Kondo effect in topological insulators,\cite{Tran_TI_Kondo_PRB10} we have no clear indication of how high the Kondo temperature is expected to be, and the nontrivial effect of spin-momentum locking on Kondo singlet formation has not been determined. These factors will be considered in a forthcoming paper.

\section{Kinetic equation}

The Liouville equation for the density operator $\hat\rho$ is projected onto the basis $\{ \ket{{\bm k}} \}$ as in Ref.~\onlinecite{Culcer_TI_Kineq_PRB10}. The density matrix $\rho_{{\bm k}{\bm k}'} = f_{{\bm k}} \, \delta_{{\bm k}{\bm k}'} + g_{{\bm k}{\bm k}'}$, where $g_{{\bm k}{\bm k}'}$ is off-diagonal in ${\bm k}$, and \textit{all} quantities are matrices in spin space. To first order in ${\bm E}$, the diagonal part $f_{\bm k}$ satisfies \cite{Culcer_TI_Kineq_PRB10}
\begin{equation}
\label{eq:f} 
\td{f_{{\bm k}}}{t} + \frac{i}{\hbar} \, [H_{0{\bm k}}, f_{{\bm k}}] + \hat{J}(f_{\bm k}) = \mathcal{D}_{\bm k}. 
\end{equation}
The scattering term $\hat{J}(f_{\bm k})$ is given below, the driving term $\mathcal{D}_{\bm k} = - \frac{i}{\hbar} \, [H^E_{\bm k}, f_{0{\bm k}}]$, and $f_{0{\bm k}}$ is the equilibrium density matrix, which is diagonal in ${\bm k}$. We write $f_{\bm k} = n_{\bm k}\openone + S_{\bm k}$, where $S_{\bm k}$ is a $2 \times 2$ Hermitian matrix. Every matrix in this problem can be written in terms of a scalar part, labeled by the subscript $n$, and ${\bm \sigma}$. Rather than choosing Cartesian coordinates to express the latter, it is more natural to identify three orthogonal directions in reciprocal space, which we denote by $\hat{\bm \Omega}_{\bm k}$, $\hat{\bm k}$, and $\hat{\bm z}_{eff}$, and project ${\bm \sigma}$ onto them. The three orthogonal directions in reciprocal space that we use for reference are
\begin{equation}
\arraycolsep 0.3ex
\begin{array}{rl}
\displaystyle \hat{\bm \Omega}_{\bm k} = & \displaystyle - a_k \, \hat{\bm \theta} + b_k \, \hat{\bm z} \\ [3ex] 
\displaystyle \hat{\bm k}_{eff} = & \displaystyle \hat{\bm k} \\ [3ex] 
\displaystyle \hat{\bm z}_{eff} = & \displaystyle a_k \, \hat{\bm z} + b_k \, \hat{\bm \theta}.
\end{array}
\end{equation}
We project ${\bm \sigma}$ as $\sigma_{{\bm k}, \parallel} = {\bm \sigma} \cdot \hat{\bm \Omega}_{\bm k}$, $\sigma_{{\bm k}, k} = {\bm \sigma} \cdot \hat{\bm k}$, and $\sigma_{{\bm k}, z_{eff}} = {\bm \sigma} \cdot \hat{\bm z}_{eff}$. Note that $\sigma_{{\bm k}, \parallel}$ commutes with $H_{0{\bm k}}$, while $\sigma_{{\bm k}, k}$ and $\sigma_{{\bm k}, z_{eff}}$ do not. Therefore we shall often use the abbreviation $\perp$ to refer to vectors in the plane spanned by $\hat{\bm k}$ and $\hat{\bm z}_{eff}$. We project $S_{\bm k}$ onto the three directions in ${\bm k}$-space, writing $S_{\bm k} = S_{{\bm k}, \parallel} + S_{{\bm k}, k} + S_{{\bm k}, z_{eff}}$, and defining $S_{{\bm k}, \parallel} = (1/2) \, s_{{\bm k}, \parallel} \, \sigma_{{\bm k}, \parallel}$, $S_{{\bm k}, k} = (1/2) \, s_{{\bm k}, k} \, \sigma_{{\bm k}, k}$ and $S_{{\bm k}, z_{eff}} = (1/2) \, s_{{\bm k}, z_{eff}} \, \sigma_{{\bm k}, z_{eff}}$. 

Equation (\ref{eq:f}) can be written as
\begin{subequations}\label{eq:Spp}
\begin{eqnarray}
\td{n_{\bm k}}{t} + P_n \hat J (f_{{\bm k}}) & = & \mathcal{D}_{{\bm k},n}, \\ [0.5ex]
\td{S_{{\bm k}, \parallel}}{t} + P_\| \hat J (f_{{\bm k}}) & = & \mathcal{D}_{{\bm k},\parallel}, \\ [0.5ex]
\label{Sperp}
\td{S_{{\bm k}, \perp}}{t} + \frac{i}{\hbar} \, [H_{{\bm k}}, S_{{\bm k}\perp}] + P_\perp \hat J (f_{{\bm k}}) & = & \mathcal{D}_{{\bm k},\perp}.
\end{eqnarray}
\end{subequations}
The projector $P_\parallel$ acts on a matrix $\mathcal{M}$ as ${\rm tr} \, (\mathcal{M}\sigma_{{\bm k}, \parallel})$, with tr the matrix trace, while $P_\perp$ singles out the part orthogonal to $H_{0{\bm k}}$. The total scattering term $\hat{J} (f_{\bm k}) = \hat{J}^{Born}(f_{\bm k}) + \hat{J}^{3rd}(f_{\bm k})$, with
\begin{equation}
\label{JBorn}
\hat{J}^{Born}(f_{\bm k}) = \bkt{\int_0^{\infty} \frac{dt'}{\hbar^2} \, [\hat U, e^{- \frac{i \hat H t'}{\hbar}}[\hat U, \hat f]\, e^{ \frac{i \hat H t'}{\hbar}}] }_{{\bm k}{\bm k}},
\end{equation}
Considering the negligible size of $b_k$ at $k = k_F$, the condition yielding the suppression of backscattering is effectively unmodified from the non-magnetic case. In the absence of scalar terms in the Hamiltonian the scalar and spin-dependent parts of the density matrix, $n_{\bm k}$ and $S_{\bm k}$, are decoupled in the first Born approximation. In the second Born approximation we obtain the additional scattering term
\begin{widetext}
\begin{equation}
\label{Jss3rd}
\hat{J}^{3rd}(f_{\bm k}) = - \frac{i}{\hbar^3} \bkt{\int_0^{\infty} dt' \int_0^{\infty} dt'' [\hat U, e^{- \frac{i \hat H t'}{\hbar}}[\hat U, e^{- \frac{i \hat H t''}{\hbar}}[\hat U, \hat f]\, e^{\frac{i \hat H t''}{\hbar}} ]\, e^{\frac{i \hat H t'}{\hbar}}]}_{{\bm k}{\bm k}},
\end{equation}
\end{widetext}
while $\bkt{}$ represents averaging over impurity configurations. The former can be further broken down into $\hat{J}^{Born}(f_{\bm k}) = \hat{J}_0(f_{\bm k}) + \hat{J}^{Born}_{ss}(f_{\bm k}) + \hat{J}^{Born}_{sj}(f_{\bm k})$. The first term, $\hat J_0(f_{{\bm k}})$, in which $\lambda = {\bm E} = 0$ in the time evolution operator, represents elastic, spin-independent, pure momentum scattering. In $\hat{J}^{Born}_{ss}(f_{\bm k})$ we allow $\lambda$ to be finite but ${\bm E} = 0$. In $\hat{J}^{Born}_{sj}(f_{\bm k})$ both $\lambda$ and ${\bm E}$ are finite, thus $\hat{J}^{Born}_{sj}(f_{\bm k})$ acts on the equilibrium density matrix $f_{0{\bm k}}$. Beyond the Born approximation we retain the leading term $\hat{J}^{3rd}(f_{\bm k}) \equiv \hat{J}^{3rd}_{ss}(f_{\bm k})$, with $\lambda$ finite but ${\bm E} = 0$, and which is customarily responsible for skew scattering \cite{Sinitsyn_AHE_Review_JPCM08}. The contribution due to magnetic impurities is also $\propto \sigma_z$ and is contained in $\hat{J}^{Born}_{ss}(f_{\bm k})$.

The formalism we adopt has a quantum mechanical foundation and accounts for the three mechanisms known to be important in the AHE. The interplay of band structure spin-orbit interaction, adiabatic change in ${\bm k}$ in the external electric field, and out-of-plane magnetization leads to a sideways displacement of carriers even in the absence of scattering (\textit{intrinsic}) \cite{Luttinger_AHE_PR58}. The spin dependence of the impurity potentials via exchange ($J$) and extrinsic spin-orbit coupling causes asymmetric scattering of up and down spins (skew scattering) \cite{Smit_SS_58}, and a sideways displacement during scattering (side jump) \cite{Berger_SJ_PRB1970}. 

\begin{widetext}

\subsection{Spin-independent scattering term in the Born approximation} 

The projections of $\hat{J}_0(f_{{\bm k}})$ that we will need in this work, derived in an analogous manner to Ref.\ \onlinecite{Culcer_TI_Kineq_PRB10}, are
\begin{equation}\label{Jproj}
\arraycolsep 0.3ex
\begin{array}{rl}
\displaystyle P_\parallel \hat{J}_0(S_{{\bm k}\parallel}) = & \displaystyle \frac{\pi n_i}{2\hbar } \int \frac{d^2k'}{(2\pi)^2} \, |U_{{\bm k}{\bm k}'}|^2 \, (s_{{\bm k}\parallel} - s_{{\bm
k}'\parallel}) \, (1 + b_k^2 + a_k^2\cos\gamma) \, \sigma_{{\bm k} \parallel} \, \delta(\varepsilon'_+ - \varepsilon_+) \\ [3ex]
\displaystyle P_\perp \hat{J}_0(S_{{\bm k}\parallel}) = & \displaystyle - \frac{\pi n_i}{2\hbar} \int \frac{d^2k'}{(2\pi)^2} \, |U_{{\bm k}{\bm k}'}|^2 (s_{{\bm k}\parallel} - s_{{\bm k}'\parallel}) \, [ (\hat{\bm \Omega}_{{\bm k}'} \cdot \hat{\bm k}) \, \sigma_{{\bm k}, k} + (\hat{\bm \Omega}_{{\bm k}'} \cdot \hat{\bm z}_{eff}) \, \sigma_{{\bm k}, z_{eff}}] \, \delta(\varepsilon'_+ - \varepsilon_+) \\ [3ex] 
\displaystyle P_\parallel \hat{J}_0(S_{{\bm k}\perp}) = & \displaystyle - \frac {\pi n_i}{2\hbar} \int \frac{d^2k'}{(2\pi)^2} \, |U_{{\bm k}{\bm k}'}|^2 [(s_{k, eff} + s'_{k,eff}) (\hat{\bm \Omega}_{{\bm k}'}\cdot\hat{\bm k}) +  ( s_{z,eff} - s'_{z,eff} ) \, (\hat{\bm \Omega}_{{\bm k}'} \cdot \hat{\bm z}_{eff}) ] \, \sigma_{{\bm k}\parallel} \, \delta(\varepsilon'_+ - \varepsilon_+),
\end{array}
\end{equation}
where $n_i$ is the impurity density, which emerges once one averages over impurity configurations. 

\subsection{Born approximation skew scattering and magnetic impurity scattering} 

The effect of this term on the scalar part of the density matrix gives a scalar contribution, which does not yield an electrical current. Next, $\hat{J}^{Born}_{ss}$ acting on the spin-dependent part of the density matrix gives
\begin{equation}
\arraycolsep 0.3ex
\begin{array}{rl}
\displaystyle \hat{J}^{Born}_{ss} (S_{\bm k}) = & \displaystyle - \frac{in_i\lambda k^2}{\hbar^2} \int\frac{d^2k'}{(2\pi)^2} \, |\mathcal{U}_{{\bm k}{\bm k}'}|^2 \sin\gamma \big[ \sigma_z e^{- i \hat H_{{\bm k}'} t'} (S_{{\bm k}} - S_{{\bm k}'}) e^{ i H_{{\bm k}} t'} - e^{- i \hat H_{{\bm k}'} t'} (\sigma_z S_{{\bm k}} - S_{{\bm k}'} \sigma_z)e^{ i H_{{\bm k}} t'} \big] + h.c.
\end{array}
\end{equation}
In this subsection we omit factors of $\hbar$ in the time evolution operators, which are restored in the final results. The two terms in the square brackets are divided as $\hat{J}^{Born}_{ss} (S_{\bm k}) = \hat{J}^{Born}_{ss,1} (S_{\bm k}) + \hat{J}^{Born}_{ss,2} (S_{\bm k})$. The first yields 
\begin{equation}
\arraycolsep 0.3ex
\begin{array}{rl}
\displaystyle \hat{J}^{Born}_{ss,1} (S_{\bm k}) = & \displaystyle \frac{\pi n_i \lambda k^2}{2\hbar} \, \sigma_z \int\frac{d^2k'}{(2\pi)^2} \, |\mathcal{U}_{{\bm k}{\bm k}'}|^2 \sin\gamma \, ({\bm S}_{{\bm k}} - {\bm S}_{{\bm k}'}) \cdot (\hat{\bm \Omega}_{\bm k} \times \hat{\bm\Omega}_{{\bm k}'}) \, \delta(\varepsilon'_+ - \varepsilon_+).
\end{array}
\end{equation}
This scattering term becomes a source term in the kinetic equation. One will only obtain a contribution linear in $\tau$ if one feeds ${\bm S}_{{\bm k}\parallel}$ into the RHS, since ${\bm S}_{{\bm k}\parallel} \propto \tau$. But at the same time ${\bm S}_{{\bm k}\parallel} \propto \hat{\bm \Omega}_{\bm k}$, and its contribution vanishes. The surviving scattering term is $\hat{J}^{Born}_{ss,2} (S_{\bm k})$, and we refer to this term simply as $\hat{J}_{ss}^{Born}(S_{\bm k})$. Its parallel projection is
\begin{equation}
\arraycolsep 0.3ex
\begin{array}{rl}
\displaystyle P_\parallel \hat{J}_{ss}^{Born}(S_{\bm k}) = & \displaystyle \frac{\pi n_i \lambda k^2 b_k}{2\hbar} \, \sigma_{{\bm k}, \parallel} \int\frac{d^2k'}{(2\pi)^2} \, |\mathcal{U}_{{\bm k}{\bm k}'}|^2 \sin\gamma \, a_k \, [\sin\gamma (s_{{\bm k}z, eff} - s_{{\bm k}'z, eff}) - b_k \, (1 - \cos\gamma) \, (s_k - s'_k)] \, \delta(\varepsilon'_+ - \varepsilon_+) \\ [3ex]
- & \displaystyle \frac{\pi n_i \lambda k^2 a_k}{2\hbar} \, \sigma_{{\bm k}, \parallel} \int\frac{d^2k'}{(2\pi)^2} \, |\mathcal{U}_{{\bm k}{\bm k}'}|^2 \sin\gamma\, \{ (s_{k, eff} + s_{k', eff}) (1 + \cos\gamma) \\ [3ex]
- & \displaystyle \sin \gamma \, [ a_k (s_{{\bm k}\parallel} - s_{{\bm k}'\parallel}) - b_k (s_{{\bm k}z, eff} - s_{{\bm k}'z, eff})] \} \, \delta(\varepsilon'_+ - \varepsilon_+).
\end{array}
\end{equation}
It does not contribute to the Hall current. Its perpendicular projections are
\begin{equation}
\arraycolsep 0.3ex
\begin{array}{rl}
\displaystyle P_k \hat{J}_{ss}^{Born}(S_{\bm k}) = & \displaystyle \frac{\pi n_i \lambda k^2 a_k}{2\hbar} \, \sigma_{{\bm k}, k} \int\frac{d^2k'}{(2\pi)^2} \, |\mathcal{U}_{{\bm k}{\bm k}'}|^2 \sin\gamma \, [a_k \sin\gamma \, (s_k + s'_k) - \, (1 - \cos\gamma) \, (s_{{\bm k}\parallel} - s_{{\bm k}'\parallel}) ] \, \delta(\varepsilon'_+ - \varepsilon_+) \\ [3ex]
\displaystyle P_{z,eff} \hat{J}_{ss}^{Born}(S_{\bm k}) = & \displaystyle \frac{\pi n_i \lambda k^2 a_k^2}{2\hbar} \, \sigma_{{\bm k}, z_{eff}} \int\frac{d^2k'}{(2\pi)^2} \, |\mathcal{U}_{{\bm k}{\bm k}'}|^2 \sin\gamma \, [ 2 \, \sin\gamma \, (s_{z, eff} - s_{z, eff'}) - b_k (1 - \cos\gamma) \, (s_k - s'_k)] \, \delta(\varepsilon'_+ - \varepsilon_+) \\ [3ex]
+ & \displaystyle \frac{\pi n_i \lambda k^2 a_k}{2\hbar} \, \sigma_{{\bm k}, z_{eff}} \int\frac{d^2k'}{(2\pi)^2} \, |\mathcal{U}_{{\bm k}{\bm k}'}|^2 \sin^2\gamma \,  b_k \, (s_{{\bm k}\parallel} - s_{{\bm k}'\parallel}) \, \delta(\varepsilon'_+ - \varepsilon_+). 
\end{array}
\end{equation}
Next, in exactly the same manner as above, taking into account the magnetic term in $\bar{U}_{{\bm k}{\bm k}'} \propto J$, we obtain the scattering term due to magnetic impurities to linear order in $b_F$ (i.e. linear in $J$), referred to as $\hat{J}^{Born}_{mag} (f_{\bm k})$. We find that $\hat{J}^{Born}_{mag} (S_{\bm k})$ is a scalar and therefore does not contribute to transport, while 
\begin{equation}\label{JBornmag}
\arraycolsep 0.3ex
\begin{array}{rl}
\displaystyle \hat{J}^{Born}_{mag} (n_{\bm k}) = & \displaystyle - \frac{n_{mag} Js a_k k \, \bar{\mathcal U}}{2\hbar^2 A}\, \frac{eE \tau \cos\theta}{2\hbar} \, \pd{f_{0+}}{k} \, {\bm \sigma}\cdot \hat{\bm z}_{eff},
\end{array}
\end{equation}
where $\bar{\mathcal U} = (1/2\pi) \int \, d\gamma \, \mathcal{U}_{{\bm k}{\bm k}'}(\gamma)$ and $n_{mag} \le n_i$. 

\subsection{Skew scattering beyond the Born approximation} 

The skew scattering term of third order in the scattering potential is 
\begin{equation}
\arraycolsep 0.3ex
\begin{array}{rl}
\displaystyle \hat{J}^{3rd}_{ss}(f_{\bm k}) = & \displaystyle - \frac{i}{\hbar^3} \, \bkt{\int_0^{\infty} dt'\, e^{- \eta t'} \int_{0}^{\infty} dt'' \, e^{-\eta t''} \, [\hat U, e^{- i \hat H_0 t'}[\hat U, e^{- i \hat H_0 t''}[\hat U, \hat f]\, e^{i \hat H_0 t''} ]\, e^{i \hat H_0 t'}]_{{\bm k}{\bm k}}}.
\end{array}
\end{equation}
In each term the $\lambda$-dependent part can appear in three places, so we have a total of 12 terms to evaluate. In each term in $[]_{{\bm k}{\bm k}}$ we retain the imaginary, spin-dependent part. We find that only $\hat{J}^{3rd}_{ss}(n_{\bm k})$ contributes to skew scattering, so we only retain this term to simplify the algebra below. We let $k_1, k_2 \rightarrow k^2$ since this term will eventually be evaluated at the Fermi energy. Skipping a large amount of laborious, yet straightforward, detail, the parallel projection of the third order skew scattering term is
\begin{equation}
\arraycolsep 0.3ex
\begin{array}{rl}
\displaystyle P_\parallel \hat{J}_{ss}^{3rd} = \frac{\lambda k^2 b_k}{4 \hbar}\, |\mathcal{U}|^3 \, (n_{{\bm k}_2} - n_{{\bm k}_1}) \, ({\bm \sigma} \cdot \hat{\bm \Omega}_{\bm k}) \, [ & \displaystyle (9 + 3b_k^2 - a_k^2) (\sin\gamma_{{\bm k}{\bm k}_1} + \sin\gamma_{{\bm k}_1{\bm k}_2} + \sin\gamma_{{\bm k}_2{\bm k}}) \\ [3ex]
- & \displaystyle \frac{3a_k^2}{2} \, (\sin2\gamma_{{\bm k}{\bm k}_1} + \sin2\gamma_{{\bm k}_1{\bm k}_2} + \sin2\gamma_{{\bm k}_2{\bm k}})]\, D_{{\bm k}_1{\bm k}} D_{{\bm k}_2{\bm k}}.
\end{array}
\end{equation}
This term vanishes when we integrate over ${\bm k}_1$ and ${\bm k}_2$, thus there is no skew scattering contribution to the AHE that is $\propto \tau$. The projections $\parallel \hat{\bm k}$ can be written as
\begin{equation}
\arraycolsep 0.3ex
\begin{array}{rl}
\displaystyle P_k \hat{J}_{ss}^{3rd, A} = & \displaystyle \frac{2a_kb_k \lambda k^2}{\hbar^3} \, |\mathcal{U}|^3 \, (n_{{\bm k}_2} - n_{\bm k}) \,  ({\bm\sigma} \cdot \hat{\bm k}) \, \big[ - \sin^2\gamma_{{\bm k}{\bm k}_1} + \sin\gamma_{{\bm k}_1{\bm k}_2} \sin\gamma_{{\bm k}{\bm k}_1} - \sin\gamma_{{\bm k}{\bm k}_1}\sin\gamma_{{\bm k}{\bm k}_2}  \big] \\ [3ex]

+ & \displaystyle \frac{2a_kb_k \lambda k^2}{\hbar^3} \, |\mathcal{U}|^3 \, (n_{{\bm k}_1} - n_{\bm k}) \,  ({\bm\sigma} \cdot \hat{\bm k}) \, \big[ - \sin^2\gamma_{{\bm k}{\bm k}_2}  - \sin\gamma_{{\bm k}_1{\bm k}_2} \sin\gamma_{{\bm k}{\bm k}_2} - \sin\gamma_{{\bm k}{\bm k}_1} \sin\gamma_{{\bm k}{\bm k}_2} \big] \\ [3ex]

\displaystyle P_k \hat{J}_{ss}^{3rd, B} = & \displaystyle \frac{2a_kb_k \lambda k^2}{\hbar^3} \, |\mathcal{U}|^3 \, (n_{{\bm k}_2} - n_{{\bm k}_1}) \,  ({\bm\sigma} \cdot \hat{\bm k}) \, \big( - \sin^2\gamma_{{\bm k}{\bm k}_1} + \sin\gamma_{{\bm k}_1{\bm k}_2} \sin\gamma_{{\bm k}{\bm k}_1} - \sin\gamma_{{\bm k}{\bm k}_1}\sin\gamma_{{\bm k}{\bm k}_2} \big) \\ [3ex]

+ & \displaystyle \frac{2a_kb_k \lambda k^2}{\hbar^3} \, |\mathcal{U}|^3 \, (n_{{\bm k}_2} - n_{{\bm k}_1}) \,  ({\bm\sigma} \cdot \hat{\bm k}) \, \big( \sin^2\gamma_{{\bm k}{\bm k}_2} + \sin\gamma_{{\bm k}_1{\bm k}_2} \sin\gamma_{{\bm k}{\bm k}_2} + \sin\gamma_{{\bm k}{\bm k}_1}\sin\gamma_{{\bm k}{\bm k}_2}  \big).
\end{array}
\end{equation}
$P_k \hat{J}_{ss}$ vanishes for isotropic scattering and does not contribute to the Hall current. The projections $\parallel \hat{\bm z}_{eff}$ can be written as
\begin{equation}
\arraycolsep 0.3ex
\begin{array}{rl}
\displaystyle P_{z,eff} \hat{J}_{ss}^{3rd, A} = \frac{\lambda k^2a_k}{\hbar^3} \, |\mathcal{U}|^3 \, (n_{{\bm k}_2} - n_{\bm k}) \, ({\bm \sigma} \cdot \hat{\bm z}_{eff}) \, \big\{ & \displaystyle \sin\gamma_{{\bm k}{\bm k}_1} [(1 + b_k^2)(1 + \cos\gamma_{{\bm k}{\bm k}_1}) + a_k^2\cos\gamma_{{\bm k}{\bm k}_2} + a_k^2\cos\gamma_{{\bm k}_1{\bm k}_2}] \\ [3ex]

+ & \displaystyle \sin\gamma_{{\bm k}_1{\bm k}_2}  [(1 + b_k^2)(1 - \cos\gamma_{{\bm k}{\bm k}_1}) + a_k^2\cos\gamma_{{\bm k}{\bm k}_2} - a_k^2\cos\gamma_{{\bm k}_1{\bm k}_2}]  \\ [3ex]

+ & \displaystyle \sin\gamma_{{\bm k}_2{\bm k}}  [(1 + b_k^2)(1 - \cos\gamma_{{\bm k}{\bm k}_1}) - a_k^2\cos\gamma_{{\bm k}{\bm k}_2} + a_k^2\cos\gamma_{{\bm k}_1{\bm k}_2}]  \big\} + ({\bm k}_1 \leftrightarrow {\bm k}_2) \\ [3ex]

\displaystyle P_{z, eff} \hat{J}_{ss}^{3rd, B} = \frac{\lambda k^2a_k}{\hbar^3} \, |\mathcal{U}|^3 \, (n_{{\bm k}_2} - n_{{\bm k}_1}) \, ({\bm \sigma} \cdot \hat{\bm z}_{eff}) \, \big\{ & \displaystyle \sin\gamma_{{\bm k}{\bm k}_1} [(1 + b_k^2)(1 + \cos\gamma_{{\bm k}{\bm k}_1}) + a_k^2\cos\gamma_{{\bm k}{\bm k}_2} + a_k^2\cos\gamma_{{\bm k}_1{\bm k}_2}] \\ [3ex]

+ & \displaystyle \sin\gamma_{{\bm k}_1{\bm k}_2}  [(1 + b_k^2)(1 - \cos\gamma_{{\bm k}{\bm k}_1}) + a_k^2\cos\gamma_{{\bm k}{\bm k}_2} - a_k^2\cos\gamma_{{\bm k}_1{\bm k}_2}]  \\ [3ex]

+ & \displaystyle \sin\gamma_{{\bm k}_2{\bm k}}  [(1 + b_k^2)(1 - \cos\gamma_{{\bm k}{\bm k}_1}) - a_k^2\cos\gamma_{{\bm k}{\bm k}_2} + a_k^2\cos\gamma_{{\bm k}_1{\bm k}_2}]  \big\} + ({\bm k}_1 \leftrightarrow {\bm k}_2).
\end{array}
\end{equation}
\end{widetext}
We will leave the results in this form until Section IV, where they will be evaluated explicitly.

\subsection{Side jump} 

One contribution due to side jump arises from the intrinsic side-jump driving term, considered in Section IV. The other two contributions come from corrections to the scattering term. We retain only the parallel projection of the side-jump scattering terms, the others being of higher order in $\hbar/(\varepsilon_F\tau)$. The easiest way is to write the kinetic equation in terms of the ${\bm k}-$diagonal and off-diagonal components of the density matrix
\begin{subequations}
\begin{eqnarray}
\label{eq:f} \td{f_{{\bm k}}}{t} + \frac{i}{\hbar} \, [H_{0{\bm k}}, f_{{\bm k}}] & = & - \frac{i}{\hbar} \, [H^E_{\bm k}, f_{{\bm k}}] - \frac{i}{\hbar} \, [\hat U, \hat g]_{{\bm k}{\bm k}} , \\ [1ex]
\label{eq:g} \td{g_{{\bm k}{\bm k}'}}{t} + \frac{i}{\hbar} \, [\hat{H}, \hat g]_{{\bm k}{\bm k}'} & = & - \frac{i}{\hbar} \, [\hat U, \hat f + \hat g]_{{\bm k}{\bm k}'}.
\end{eqnarray}
\end{subequations}
In the Born approximation, we solve Eq.\ (\ref{eq:g}) to first order in $\hat{U}$
\begin{equation}\label{eq:gsol}
g_{{\bm k}{\bm k}'} = - \frac{i}{\hbar} \, \int_0^\infty dt'\, e^{- i \hat H t'} \left[\hat U, \hat f(t - t') \right] e^{i \hat H t'}\bigg|_{{\bm k}{\bm k}'}.
\end{equation}
To include side jump, we take into account an additional correction to $g$ arising from $H_E$. Still in the Born approximation we write $g = g_0 + g_E$
\begin{widetext}
\begin{equation}
\arraycolsep 0.3ex
\begin{array}{rl}
\displaystyle \td{g_{0, {\bm k}{\bm k}'}}{t} + \frac{i}{\hbar} \, [\hat{H}_0, \hat{g}_0]_{{\bm k}{\bm k}'} & \displaystyle = - \frac{i}{\hbar} \, [\hat U, \hat f]_{{\bm k}{\bm
k}'} \\ [3ex]
\displaystyle g_{0, {\bm k}{\bm k}'} = & \displaystyle - \frac{i}{\hbar} \, \bigg\{ \int_0^\infty dt'\, e^{-\eta t'}\,  e^{- i \hat{H}_0 t'}[\hat U, \hat f(t - t')]\, e^{i \hat{H}_0 t'}\bigg\}_{{\bm k}{\bm k}'} \\ [3ex]
\displaystyle \td{g_{E, {\bm k}{\bm k}'}}{t} + \frac{i}{\hbar} \, [\hat{H}_0, \hat{g}_E]_{{\bm k}{\bm k}'} & \displaystyle = - \frac{i}{\hbar} \, [\hat{H}_E, \hat{g}_0]_{{\bm k}{\bm
k}'} \\ [3ex]
\displaystyle g_{E, {\bm k}{\bm k}'} = & \displaystyle - \frac{i}{\hbar} \, \bigg\{ \int_0^\infty dt'\, e^{-\eta t'}\,  e^{- i \hat{H}_0 t'}[\hat{H}_E, \hat{g}_0(t - t')]\, e^{i \hat{H}_0 t'}\bigg\}_{{\bm k}{\bm k}'}.
\end{array}
\end{equation}
Factors of $\hbar$ have been left out of the time evolution operators to shorten the equations. We adopt the Markovian approximation $\hat{f}(t - t') \approx \hat{f}(t) \equiv \hat{f}$, and, remembering that $\hat{H}_E$ is first order in the electric field, we replace $\hat{f} \rightarrow \hat{f}_0$, the equilibrium density operator
\begin{equation}
\arraycolsep 0.3ex
\begin{array}{rl}
\displaystyle g_{E, {\bm k}{\bm k}'} (t) \approx & \displaystyle - \frac{1}{\hbar^2} \, \bigg\{ \int_0^\infty dt''\, e^{-\eta t''} e^{- i \hat{H}_0 t''}[\hat{H}_E, e^{- i \hat{H}_0 t'}[ \hat U, \hat{f}_0 ]\, e^{i \hat{H}_0 t'}] \, e^{i \hat{H}_0 t''}\bigg\}_{{\bm k}{\bm k}'}.
\end{array}
\end{equation}
Both $\hat{H}_E$ and $\hat{U}$ have scalar and spin-orbit parts. We first evaluate the easier contribution from $\displaystyle H_{E, {\bm k}{\bm k}'}^{sj}$ and the scalar part of the scattering potential, which we call $g_{E, {\bm k}_1{\bm k}_2}^{sj}$. This contribution can be expressed as
\begin{equation}
\arraycolsep 0.3ex
\begin{array}{rl}
\displaystyle g_{E, {\bm k}_1{\bm k}_2}^{sj} = & \displaystyle - \frac{i}{\hbar} \, \int_0^\infty dt'\, e^{-\eta t'}\,  e^{- i {H}_{0{\bm k}_1} t'}(H_{E{\bm k}_1}^{sj} g_{0{\bm k}_1{\bm k}_2} - g_{0{\bm k}_1{\bm k}_2} H_{E{\bm k}_2}^{sj})\, e^{i H_{0{\bm k}_2} t'}.
\end{array}
\end{equation}
where $g_{0{\bm k}_1{\bm k}_2}$ is taken to zeroth order in $\lambda$. This gives rise to the scattering term
\begin{equation}\label{Jsj}
\arraycolsep 0.3ex
\begin{array}{rl}
\displaystyle \hat{J}_{sj} (f_{0{\bm k}}) = & \displaystyle \frac{i}{\hbar} \sum_{{\bm k}_1} \bar{U}_{{\bm k}{\bm k}_1}g_{E, {\bm k}_1{\bm k}}^{sj} + h.c,
\end{array}
\end{equation}
where only the scalar part of $\bar{U}_{{\bm k}{\bm k}_1}$ contributes, since $g_{E, {\bm k}_1{\bm k}}^{sj}$ is already first order in $\lambda$.

We require in addition the scattering term
\begin{equation}\label{Jsc}
\arraycolsep 0.3ex
\begin{array}{rl}
\displaystyle \hat{J}_{sc} (f_{0{\bm k}}) = & \displaystyle \frac{i}{\hbar} \sum_{{\bm k}_1} \bar{U}_{{\bm k}{\bm k}_1}g_{E, {\bm k}_1{\bm k}}^{sc} + h.c.
\end{array}
\end{equation}
The term $\hat{J}_{sc}(f_{0{\bm k}}) $ has two contributions. One contribution arises from the spin-independent part of $|\bar{U}_{{\bm k}{\bm k}'}|$ and $g_0$ to first order in $\lambda$. The other term comes from the spin-dependent part of $|\bar{U}_{{\bm k}{\bm k}'}|$ and $g_0$ to zeroth order in $\lambda$. Therefore we only need to evaluate one other term to first order in $\lambda$,
\begin{equation}
\arraycolsep 0.3ex
\begin{array}{rl}
\displaystyle g_{E, {\bm k}_1{\bm k}_3}^{sc} = & \displaystyle - \frac{i}{\hbar} \, \int_0^\infty dt'\, e^{-\eta t'}\,  e^{- i {H}_{0{\bm k}_1} t'}(H_{E{\bm k}_1{\bm k}_2}^{sc} g_{0{\bm k}_2{\bm k}_3} - g_{0{\bm k}_1{\bm k}_2}H_{E{\bm k}_2{\bm k}_3}^{sc})\, e^{i H_{0{\bm k}_3} t'}.
\end{array}
\end{equation}
We will leave the result in this form until Section IV when it will be evaluated explicitly.
\end{widetext}

\section{Anomalous Hall effect}

We solve Eq.\ (\ref{eq:Spp}) perturbatively in the small parameter $\hbar/(\varepsilon_F\tau)$ (which is proportional to the impurity density $n_i$) as described in Ref. \onlinecite{Culcer_TI_Kineq_PRB10}, as well as in $\lambda$. In transport this expansion starts at order $[\hbar/(\varepsilon_F\tau)]^{-1}$ (i.e. $n_i^{-1}$), reflecting the competition between the driving electric field and impurity scattering resulting in a shift of the Fermi surface. The leading terms in $[\hbar/(\varepsilon_F\tau)]^{-1}$ appear in $n_{\bm k}$ and $S_{{\bm k}, \parallel}$ \cite{Culcer_TI_Kineq_PRB10}, namely in the components of the density matrix which commute with $H_{0{\bm k}}$. Identifying these terms correctly is crucial, since if terms $\propto n_i^{-1}$ exist in the AHE they will be dominant at high mobilities.

We first solve Eqs.\ (\ref{eq:Spp}) to zeroth order in $\lambda$, setting $\hat J (f_{{\bm k}}) \rightarrow \hat{J}_0 (f_{{\bm k}})$. We refer to this part of the solution as the \textit{bare} part, which is gives the intrinsic contribution to the AHE (i.e. $\lambda = 0$.) Once the bare components of $S_{{\bm k}}$ are determined, we project repeatedly between $S_{{\bm k}, \parallel}$ and $S_{{\bm k}, \perp}$ in Eq.\ (\ref{eq:Spp}) until we have found all terms to $[\hbar/(\varepsilon_F\tau)]^{-1}$ and $[\hbar/(\varepsilon_F\tau)]^{0}$. These projections give \textit{disorder renormalizations}, explained in a separate subsection below. Subsequently, the scattering terms $\hat{J}^{Born}_{ss}(f_{\bm k}) + \hat{J}^{3rd}(f_{\bm k})$ act on the bare solution to generate new driving terms to first order in $\lambda$, for which an analogous solution is found to the density matrix. These give the skew scattering contributions, and the contribution due to magnetic impurities. Finally the side-jump driving terms are found, in which $f_{\bm k} \rightarrow f_{0{\bm k}}$. Given the lengthy expressions, below we present only the terms that contribute to the AHE.

The appearance of terms to order zero in $\hbar/(\varepsilon_F\tau)$ is standard in transport and important in spin-dependent transport. These terms depend on the angular structure of the scattering potential, thus it is not enough to study only the band structure contribution, but also its disorder renormalization. Disorder renormalizations are crucial in spin-related transport, where they can cancel band structure contributions \cite{Inoue_RashbaSHE_Vertex_PRB04}. Though additional terms of order $[\hbar/(\varepsilon_F\tau)]^{0}$ may exist, they arise from scattering terms $\propto n_i^2$, describing quantum interference and related effects relevant at higher impurity densities.

\subsection{Bare driving term}

To first order in ${\bm E}$, we write $f_{\bm k} = f_{0{\bm k}} + f_{E{\bm k}}$, with corresponding notation for $n_{\bm k}$ and $S_{\bm k}$. The Fermi-Dirac function for the positive energy branch is denoted by $f_{0+} = f_0(\varepsilon_+)$, with $f_{0-} = 0$ for electron doping. The equilibrium density matrix is $f_{0{\bm k}} = \frac{1}{2} \, (f_{0+} + f_{0-}) \openone + \frac{1}{2} \, (f_{0+} - f_{0-}) \sigma_{{\bm k}, \parallel}$. The driving term is 
\begin{equation}
\arraycolsep 0.3ex
\begin{array}{rl}
\displaystyle \mathcal{D}_{\bm k} = & \displaystyle - \frac{i}{\hbar} \, [H^E_{\bm k}, f_{0{\bm k}}] \\ [3ex]
= & \displaystyle (e{\bm E}/\hbar) \cdot \pd{f_{0{\bm k}}}{{\bm k}} - \frac{ie\lambda}{\hbar} \, [{\bm \sigma}\cdot({\bm k}\times{\bm E}), f_{0{\bm k}}]. 
\end{array}
\end{equation}
Its projections are $\mathcal{D}_{\bm k} = \mathcal{D}_{{\bm k},n} + \mathcal{D}_{{\bm k},\parallel} + \mathcal{D}_{{\bm k}, k} + \mathcal{D}_{{\bm k}, z_{eff}}$. To zeroth order in $\lambda$ 
\begin{equation}
\arraycolsep 0.3ex
\begin{array}{rl}
\displaystyle \mathcal{D}_{{\bm k}, n} = & \displaystyle \frac{e{\bm E} \cdot \hat{\bm k}}{2\hbar} \, \bigg( \pd{f_{0+}}{k}\bigg) \\ [3ex]
\displaystyle \mathcal{D}_{{\bm k}, \parallel} = & \displaystyle \frac{1}{2}\, \sigma_{{\bm k}, \parallel} \, \frac{e{\bm E} \cdot \hat{\bm k}}{\hbar} \, \bigg( \pd{f_{0+}}{k} \bigg) \\ [3ex]
\displaystyle \mathcal{D}_{{\bm k}, k} = & \displaystyle \frac{1}{2}\, \sigma_{{\bm k}, k} \, \frac{e{\bm E} \cdot \hat{\bm \theta}}{\hbar k} \, a_k \, (f_{0+}) \\ [3ex]
\displaystyle \mathcal{D}_{{\bm k}, z_{eff}} = & \displaystyle - \frac{1}{2}\, \sigma_{{\bm k}, z_{eff}} \, \frac{e{\bm E} \cdot \hat{\bm k}}{\hbar k} \, a_k b_k \, (f_{0+}).
\end{array}
\end{equation}
Henceforth ${\bm E} \parallel \hat{\bm x}$. We obtain
\begin{equation}\label{tauWzeta}
\arraycolsep 0.3ex
\begin{array}{rl}
\displaystyle n_{E{\bm k}}^{bare} = & \displaystyle \frac{e{\bm E} \tau \cdot \hat{\bm k}}{2\hbar} \, \pd{f_{0+}}{k} \\ [3ex]
\displaystyle S_{E{\bm k}, \parallel}^{bare} = & \displaystyle \frac{1}{2}\, {\bm \sigma} \cdot \hat{\bm \Omega}_{\bm k} \, \frac{e{\bm E} \tau \cdot \hat{\bm k}}{\hbar} \, \pd{f_{0+}}{k}.
\end{array}
\end{equation}
The transport scattering time for screened Coulomb impurities is defined as
\begin{equation}
\arraycolsep 0.3ex
\begin{array}{rl}
\displaystyle \frac{1}{\tau} = & \displaystyle \frac{\varepsilon_+}{A} \, \frac{n_iW_k}{2\hbar A} \, (\zeta_0 - \zeta_1) \\ [3ex]
\displaystyle W_k = & \displaystyle \frac{Z^2e^4}{4\varepsilon_0^2\varepsilon_r^2k_F^2} \\ [3ex]
\displaystyle \zeta = & \displaystyle \frac{1}{2} \, (1 + b_k^2 + a_k^2\cos\gamma)\frac{1}{\big(\sin\frac{\gamma}{2} + \frac{k_{TF}}{2k_F} \big)^2}.
\end{array}
\end{equation}
$\zeta$ contains a factor of $1/2$ as it is defined in analogy with the ${\bm B} = 0$ case, \cite{Culcer_TI_Kineq_PRB10} while $\zeta_{0,1}$ refer to its Fourier components, and $\frac{k_{TF}}{2k_F} = \frac{r_s}{4}$, with the Wigner-Seitz radius $r_s = e^2/(2\pi \epsilon_0 \epsilon_r A)$. This way $|U_{{\bm k}{\bm k}'}|^2 = W_k/\big(\sin\frac{\gamma}{2} + \frac{r_s}{4} \big)^2$. 

The (bare) perpendicular part of the spin density matrix is
\begin{equation}
\arraycolsep 0.3ex
\begin{array}{rl}
\displaystyle S_{E{\bm k}, z_{eff}}^{bare} = & \displaystyle \frac{1}{2}\, {\bm \sigma} \cdot \hat{\bm z}_{eff} \, \frac{e{\bm E} \cdot \hat{\bm \theta}}{\hbar k \Omega}\, a_k \, f_{0+} \\ [3ex]
\displaystyle S_{E{\bm k}, k}^{bare} = & \displaystyle \frac{1}{2}\, {\bm \sigma} \cdot \hat{\bm k} \, \frac{e{\bm E} \cdot \hat{\bm k}}{\hbar k \Omega} \, a_k b_k \, f_{0+}.
\end{array}
\end{equation}
This completes the evaluation of the \textit{bare} correction to the spin density matrix, $n_{E{\bm k}}^{bare} + S_{E{\bm k}}^{bare}$, which will yield the intrinsic contribution to the Hall current and its disorder renormalization. 

\subsection{Disorder renormalizations}
\label{sec:renorm}

The correction to the bare density matrix is renormalized by scattering between $S_{{\bm k}, \perp}$ and $S_{{\bm k}, \parallel}$. In addition, a number of terms appearing below also need to be renormalized in this way. We briefly describe the technical procedure in this subsection.

Projections of the form $-P_\perp \hat J(S_{{\bm k},\parallel})$ involve no complications, since Eq.\ (\ref{Sperp}) is solved immediately by applying the time evolution operator. \cite{Culcer_TI_Kineq_PRB10} For projections of the form $-P_\parallel \hat J(S_{{\bm k}, \perp})$, it is helpful to observe that the Fourier expansions of $s_k$ and $s_{z,eff}$ throughout this work will only have Fourier component 1 (i.e. $e^{\pm i\theta}$), thus generally, using Eqs.,  
\begin{equation}\label{vertex}
\arraycolsep 0.3ex
\begin{array}{rl}
\displaystyle -P_\parallel \hat{J}_0(S_{E{\bm k}, k}) = & \displaystyle \frac{a_kb_k s_k}{\tau_{c+}} \, \sigma_{{\bm k}, \parallel} \\ [3ex]
\displaystyle -P_\parallel \hat{J}_0(S_{E{\bm k}, z_{eff}}) = & \displaystyle \frac{a_kb_ks_{z_{eff}}}{\tau_\mu} \, \sigma_{{\bm k}, \parallel}.
\end{array}
\end{equation}
The effective scattering times $\tau_\mu$, $\tau_{c+}$ are given below. The projected terms $-P_\perp \hat{J}_0(S_{{\bm k}, \parallel})$ and $-P_\parallel \hat J(S_{{\bm k}, \perp})$ then act as new driving terms for $S_{{\bm k}, \perp}$ and $S_{{\bm k}, \parallel}$ respectively, in effect renormalizing the bare results by a multiplicative constant which depends on the angular characteristics of the scattering potential. All terms in this problem have to be renormalized in this way. Yet \textit{all} terms in the density matrix consist of Fourier component 1 (that is, they are of the form $e^{\pm i\gamma}$, or alternatively $\sin\gamma$ and $\cos\gamma$), thus the renormalizations all follow the pattern of Eqs. (\ref{vertex}).

Only the bare term $S_{E{\bm k}}^{bare}$ needs to be renormalized due to scattering from $S_{{\bm k}\parallel}$ to $S_{{\bm k}\perp}$. The corrections to the perpendicular part of the spin density matrix due to scattering from $S_{E{\bm k}\parallel}^{bare}$ into $S_{{\bm k}, k}$ and $S_{{\bm k}, z_{eff}}$ are
\begin{equation}\label{SEvtx}
\arraycolsep 0.3ex
\begin{array}{rl}
\displaystyle S_{E{\bm k}, k}^{vtx} = & \displaystyle \frac{eE_x b_k \tau \cos\theta}{\hbar \tau_{c-} \Omega} \, \pd{f_{0+}}{k} \, {\bm \sigma}\cdot \hat{\bm k} \\ [3ex]
\displaystyle S_{E{\bm k}, z_{eff}}^{vtx} = & \displaystyle - \frac{eE_x \tau \sin\theta}{\hbar \tau_s \Omega} \, \pd{f_{0+}}{k} \, {\bm \sigma}\cdot \hat{\bm z}_{eff},
\end{array}
\end{equation}
The effective scattering times are
\begin{equation}
\arraycolsep 0.3ex
\begin{array}{rl}
\displaystyle \frac{1}{\tau_{s}} = & \displaystyle \frac{n_i W_k |s_1|}{2\hbar A} \, \frac{\varepsilon_+}{A} \\ [1ex]
\displaystyle \frac{1}{\tau_{c\pm}} = & \displaystyle \frac{n_i W_k}{2\hbar A} \, \frac{\varepsilon_+}{A} \, (c_0 \pm c_1) \\ [1ex]
\displaystyle \frac{1}{\tau_\mu} = & \displaystyle \frac{n_i W_k}{2\hbar A} \, \frac{\varepsilon_+}{A} \, (\mu_0 - \mu_1) \\ [1ex]
\displaystyle c (\gamma) = & \displaystyle \frac{1}{2} \, \frac{\cos\gamma}{\big(\sin\frac{\gamma}{2} + \frac{r_s}{4} \big)^2} \\ [1ex] 
\displaystyle s (\gamma) = & \displaystyle \frac{1}{2} \, \frac{\sin\gamma}{\big(\sin\frac{\gamma}{2} + \frac{r_s}{4} \big)^2} \\ [1ex]
\displaystyle \mu (\gamma) = & \displaystyle \frac{1}{2} \, \frac{1 - \cos\gamma}{\big(\sin\frac{\gamma}{2} + \frac{r_s}{4} \big)^2}.
\end{array}
\end{equation}
The renormalizations of all the other terms appearing below due to scattering from $S_{{\bm k}\parallel}$ to $S_{{\bm k}\perp}$ are of higher orders in $\lambda$ or $\hbar/(\varepsilon_F\tau)$ and need not be considered. Renormalizations due to scattering from $S_{{\bm k}\perp}$ to $S_{{\bm k}\parallel}$ can be worked out immediately from Eq.\ (\ref{Jproj}).

We emphasize that, aside from the transport relaxation time $\tau$, the effective scattering times appearing in this work are defined for convenience and they do not necessarily represent specific scattering processes, but are auxiliary quantities in achieving a final result.

\subsection{Extrinsic contributions to the density matrix}

The next step comprises two parts. Firstly, we feed $n_{E{\bm k}}^{bare} + S_{E{\bm k}}^{bare}$ into $\hat{J}^{Born}(f_{\bm k})$ and $\hat{J}^{3rd}(f_{\bm k})$, with the full spin-dependent $U_{{\bm k}{\bm k}'}$ in both, generating new source terms $\propto \lambda$ in the kinetic equation. The terms linear in $\lambda$ in $\hat{J}^{Born}(f_{E{\bm k}})$ yield
\begin{equation}
\arraycolsep 0.3ex
\begin{array}{rl}
\displaystyle S^{ss,Born}_{E{\bm k}, k} = & \displaystyle - \lambda k^2 a_k b_k \, \frac{eE_x \cos\theta}{\hbar \Omega} \, \frac{\tau}{\tau_{ss}^{Born}} \, \delta(k - k_F) \, {\bm \sigma}\cdot\hat{\bm k}.
\end{array}
\end{equation}
The effective scattering time $\tau_{ss}^{Born}$ is defined as
\begin{equation}
\arraycolsep 0.3ex
\begin{array}{rl}
\displaystyle \frac{1}{\tau_{ss}^{Born}} = & \displaystyle \frac{n_i W_k \varepsilon_+}{2 \hbar A^2} \, (\eta_0 - \eta_1) \\ [3ex]
\displaystyle \eta (\gamma) = & \displaystyle \frac{1}{2} \, \frac{\sin^2\gamma}{\big(\sin\frac{\gamma}{2} + \frac{r_s}{4} \big)^2}.
\end{array}
\end{equation}
Magnetic impurity scattering is also part of $\hat{J}^{Born}(f_{E{\bm k}})$. The scattering term $\hat{J}^{Born}_{mag} (n_{\bm k})$ evaluated in Eq. \ (\ref{JBornmag}) yields a driving term for $S_{E{\bm k}\perp}$, which gives a correction $\propto n_i\tau$ and therefore independent of the impurity density
\begin{equation}
\arraycolsep 0.3ex
\begin{array}{rl}
\displaystyle S_{E{\bm k}\perp}^{mag} = & \displaystyle a_k \,\frac{\tau}{\tau_{mag}}\, \frac{eE \cos\theta}{2\hbar \Omega_{\bm k}} \, \delta(k - k_F) \, {\bm \sigma} \cdot \hat{\bm k}.
\end{array}
\end{equation}
Here $1/\tau_{mag} = n_{mag} Js k \, \bar{\mathcal U}/(2\hbar^2 A)$. 

We come to the skew scattering term beyond the Born approximation, $\hat{J}_{ss}^{3rd}(f_{E{\bm k}})$. In the term $\hat{J}_{ss}^{3rd}(f_{E{\bm k}}^{bare})$ only $\hat{J}_{ss}^{3rd}(n_{E{\bm k}}^{bare})$ is real and finite, and we find that the projection $P_k \hat{J}_{ss}^{3rd}$ does not contribute to the AHE. We concentrate on the sum of the terms $P_{z,eff} \hat{J}_{ss}^{3rd, A} + P_{z,eff} \hat{J}_{ss}^{3rd, B} = P_{z,eff} \hat{J}_{ss}^{3rd}$. This scattering term will act on the electric-field induced correction to the scalar part of the density matrix, $n_{E{\bm k}}$, which has the angular structure $n_{E{\bm k}} = n(k) \, \cos\theta$. In light of this detail, we can boil the term $P_{z,eff} \hat{J}_{ss}^{3rd}$ to
\begin{equation}
\arraycolsep 0.3ex
\begin{array}{rl}
\displaystyle P_{z,eff} \hat{J}_{ss}^{3rd} = & \displaystyle \lambda k^2a_k n (k)  \, \frac{\sin\theta}{\tau_{ss}^{3rd}} \, ({\bm \sigma} \cdot \hat{\bm z}_{eff}) \\ [3ex]
\displaystyle \frac{1}{\tau_{ss}^{3rd}} = & \displaystyle \frac{n_i W_k}{2\hbar A} \, \frac{\varepsilon_+}{A} \, \frac{\pi r_s \phi_z^s}{4},
\end{array}
\end{equation}
where the following auxiliary quantities have been defined
\begin{widetext}
\begin{equation}
\arraycolsep 0.3ex
\begin{array}{rl}
\displaystyle \phi_z^s = & \displaystyle \frac{1}{\pi} \int d\theta\int \frac{d\theta_1}{2\pi} \int \frac{d\theta_2}{2\pi} \, \frac{F_z^s \, \sin\theta \, (2\cos\theta_2 - \cos\theta - \cos\theta_1)}{\big(\sin\frac{\gamma_{{\bm k}{\bm k}_1}}{2} + \frac{r_s}{4}\big)\big(\sin\frac{\gamma_{{\bm k}_1{\bm k}_2}}{2} + \frac{r_s}{4}\big)\big(\sin\frac{\gamma_{{\bm k}_2{\bm k}}}{2} + \frac{r_s}{4}\big)} \\ [3ex]
\displaystyle F_z^s = & \displaystyle \sin\gamma_{{\bm k}{\bm k}_1} [(1 + \cos\gamma_{{\bm k}{\bm k}_1}) + \cos\gamma_{{\bm k}{\bm k}_2} + \cos\gamma_{{\bm k}_1{\bm k}_2}] + \sin\gamma_{{\bm k}_1{\bm k}_2}  [(1 - \cos\gamma_{{\bm k}{\bm k}_1}) + \cos\gamma_{{\bm k}{\bm k}_2} - \cos\gamma_{{\bm k}_1{\bm k}_2}] \\ [3ex]
+ & \displaystyle \sin\gamma_{{\bm k}_2{\bm k}}  [(1 - \cos\gamma_{{\bm k}{\bm k}_1}) - \cos\gamma_{{\bm k}{\bm k}_2} + \cos\gamma_{{\bm k}_1{\bm k}_2}].
\end{array}
\end{equation}
\end{widetext}
The contribution of this scattering source term to the spin density matrix is
\begin{equation}
\arraycolsep 0.3ex
\begin{array}{rl}\label{Sss3rd}
\displaystyle S^{ss, 3rd}_{{\bm k}, k} = & \displaystyle \frac{\lambda k^2a_k n(k)}{\Omega} \, \frac{\sin\theta}{\tau_{ss}^{3rd}} \, ({\bm \sigma} \cdot \hat{\bm k}).
\end{array}
\end{equation}
The term $S^{ss, 3rd}_{{\bm k}, k}$ found in Eq. \ (\ref{Sss3rd}) above does not contribute to the Hall current, but its disorder renormalization yields,
\begin{equation}
\arraycolsep 0.3ex
\begin{array}{rl}
\displaystyle S^{ss, 3rd}_{E{\bm k}, \parallel} = & \displaystyle - \lambda k^2a_k^2b_k \, \frac{eE \, \sin\theta}{2\hbar\Omega_{\bm k}} \, \frac{\tau^2}{\tau_{ss}^{3rd}\tau_{c+}} \, \delta(k - k_F) \, \sigma_{{\bm k},\parallel}.
\end{array}
\end{equation}
This term contributes to the Hall current. It is the only skew scattering contribution to the anomalous Hall effect, and its value is determined explicitly below.

Finally, the side-jump scattering term $\hat{J}^{Born}_{sj} (f_{0{\bm k}})$ contains both electric field interaction terms $H_{E, {\bm k}{\bm k}'}^{sc} + H_{E, {\bm k}{\bm k}'}^{sj}$ in the time evolution operators. It is $\propto [\hbar/(\varepsilon_F\tau)]^{-1}$, thus only its parallel projection yields a term of order $[\hbar/(\varepsilon_F\tau)]^{0}$. The side-jump scattering term consists of $J_E^{sj} (f_{0{\bm k}}) + \hat{J}_{sc}(f_{0{\bm k}})$, defined in Eqs. \ (\ref{Jsj}) and (\ref{Jsc}) respectively. Substituting for $n_{E{\bm k}}$ we obtain for $P_\parallel J_E^{sj} (f_{0{\bm k}})$, after a lot of algebra,
\begin{widetext}
\begin{equation}
\arraycolsep 0.3ex
\begin{array}{rl}
\displaystyle P_\parallel J_E^{sj} (f_{0{\bm k}}) = & \displaystyle -\frac{\pi n_i e b_k \lambda |{\bm E}|}{\hbar}\, \pd{f_{0+}}{\varepsilon_+}\, \sigma_{{\bm k}, \parallel} \,  \int\frac{d^2k'}{(2\pi)^2} |\mathcal{U}_{{\bm k}{\bm k}'}|^2 \, (1 + \hat{\bm \Omega}_{{\bm k}}\cdot\hat{\bm\Omega}_{{\bm k}'} ) \, (k_y - k_y') \, \delta(\varepsilon_+' - \varepsilon_+).
\end{array}
\end{equation}
The term $\hat{J}_{sc}(f_{0{\bm k}}) $ has two contributions: one from the spin-independent part of $|\bar{U}_{{\bm k}{\bm k}'}|$ and $g_0$ to first order in $\lambda$
\begin{equation}
\arraycolsep 0.3ex
\begin{array}{rl}
\displaystyle P_\parallel \hat{J}_{sc}^1(f_{0{\bm k}})  = & \displaystyle - \frac{\pi e b\lambda |{\bm E}|}{\hbar} \, \sigma_{{\bm k}, \parallel} \, \pd{f_{0{\bm k}}}{\varepsilon_{{\bm k}}} \int\frac{d^2k'}{(2\pi)^2} \,  |\mathcal{U}_{{\bm k}{\bm k}'}|^2 \, (k_y - k_y') \, \delta(\varepsilon_{{\bm k}'} - \varepsilon_{{\bm k}}),
\end{array}
\end{equation}
and another from the spin-dependent part of $|\bar{U}_{{\bm k}{\bm k}'}|$ and $g_0$ to zeroth order in $\lambda$
\begin{equation}
\arraycolsep 0.3ex
\begin{array}{rl}
\displaystyle P_\parallel \hat{J}_{sc}^2 (f_{0{\bm k}})  = & \displaystyle \frac{\pi n_i e\lambda b k^2}{2\hbar} \, \pd{f_{0+}}{\varepsilon_+}  \, \sigma_{{\bm k}, \parallel} \int\frac{d^2k'}{(2\pi)^2} \, |\mathcal{U}_{{\bm k}{\bm k}'}|^2 \sin\gamma_{{\bm k}{\bm k}'} \, (1 + \hat{\bm \Omega}_{{\bm k}'}\cdot\hat{\bm \Omega}_{{\bm k}}) \, \bigg[ {\bm E}\cdot \pd{}{{\bm k}} \, \big(\hat{\bm \Omega}_{{\bm k}}\cdot\hat{\bm \Omega}_{{\bm k}'}\big)\bigg] \, \delta(\varepsilon_+' - \varepsilon_+).
\end{array}
\end{equation}
\end{widetext}
Integrating by parts and adding $P_\parallel \hat{J}_{sc}^1(f_{0{\bm k}})  + P_\parallel \hat{J}_{sc}^2(f_{0{\bm k}})  = P_\parallel \hat{J}_E^{sj} (f_{0{\bm k}}) $ yields an overall factor of 2 in the parallel projection of the side-jump scattering term. The overall result for the side-jump scattering contribution is
\begin{equation}\label{sjsc}
\arraycolsep 0.3ex
\begin{array}{rl}
\displaystyle S^{Jsj}_{E{\bm k}, \parallel} = - & \displaystyle \frac{4 e b_k \lambda E_x \varepsilon_+}{A^2} \, \sin\theta \, \delta(k - k_F) \,  \sigma_{{\bm k}, \parallel}.
\end{array}
\end{equation}
A further contribution due to side jump arises from the intrinsic side-jump driving term, 
\begin{equation}
\arraycolsep 0.3ex
\begin{array}{rl}
\displaystyle - \frac{i}{\hbar} \, [H_E^{sj}, f_{0{\bm k}}] = & \displaystyle - \frac{ea_kE\lambda k_y}{\hbar}\, (f_{0+} - f_{0-}) \, {\bm \sigma} \cdot \hat{\bm k}.
\end{array}
\end{equation}
This term only contributes to the perpendicular driving term and yields a correction to $S_{E{\bm k}\perp}$, found straightforwardly evaluated by applying the time evolution operator. We note in passing that this term is also responsible for the survival of side-jump spin-Hall effect in systems with band structure spin-orbit coupling \cite{Tse_SDS_IESHE_PRB06}. In the AHE in TI it yields
\begin{equation}\label{sjint}
\arraycolsep 0.3ex
\begin{array}{rl}
\displaystyle S^{sj, int}_{E{\bm k}, z_{eff}} = & \displaystyle - \frac{ea_kE_x\lambda k \sin\theta}{\hbar \Omega}\, f_{0+} \, \sigma_{{\bm k}, z_{eff}}. 
\end{array}
\end{equation}
This term is projected back onto $S_{{\bm k}\parallel}$ as described in Sec.\ \ref{sec:renorm}, giving an additional renormalization. Both Eqs.\ (\ref{sjsc}) and (\ref{sjint}) contribute to the AHE.

The current operator ${\bm j}$ has contributions from the band Hamiltonian, ${\bm j}_0 = \frac{eA}{\hbar} \,{\bm \sigma} \times \hat{\bm z}$, as well as ${\bm j}_E = \frac{2e^2 \lambda}{\hbar} \, {\bm \sigma} \times {\bm E}$, and ${\bm j}_U =  \frac{2ie\lambda}{\hbar} \, {\bm \sigma}\times ({\bm k} - {\bm k}') \, \mathcal{U}_{{\bm k }{\bm k}'}$. These latter two cancel, as they represent the force acting on the system. The central result of our work is the AHE conductivity, which, ignoring terms of order $b_F^2$, can be divided into two terms ($int\equiv$ intrinsic, $ext\equiv$ extrinsic)
\begin{equation}
\arraycolsep 0.3ex
\begin{array}{rl}
\displaystyle \sigma_{yx}^{int} = & \displaystyle - \frac{e^2}{2h} \, \bigg[1 - \frac{\pi}{8}\frac{\tau}{\tau_\mu} + b_F \tau \bigg(\frac{1}{\tau_{c-}} + \frac{1}{\tau_s} \bigg) + \frac{\tau}{2\tau_{mag}} \bigg] \\ [3ex]
\displaystyle \sigma_{yx}^{ext} = & \displaystyle \frac{e^2}{2h} \, b_F \, (\lambda k_F^2) \, \bigg(9 - \frac{8\tau}{\tau_\mu} + \frac{\tau}{\tau_{ss}^{Born}} + \frac{\tau^2}{2\tau_{ss}^{3rd}\tau_{c+}} \bigg).
\end{array}
\end{equation}
All $\tau$s are defined in the supplement, and $1/\tau_{mag} \propto b_F$, rendering this term negligible. Both $S_{{\bm k}, z_{eff}}^{bare}$ and $S_{{\bm k}, k}^{bare}$ contribute $- \frac{e^2}{4h}$ to $\sigma_{yx}^{int}$. The total ($- \frac{e^2}{2h}$) can also be expressed in terms of the Berry curvature \cite{Culcer_AHE_PRB03}, and is thus a topological quantity. It is similar to the result of Ref.~\onlinecite{ZangNagaosa_TI_Monopole_PRB10} in the vicinity of a ferromagnetic layer, and can be identified with a monopole. The corrections are disorder renormalizations due to scattering between $S_{{\bm k}, \perp}$ and $S_{{\bm k}, \parallel}$. In $\sigma_{yx}^{ext}$ the first two terms in brackets represent the combined contribution of side jump scattering and intrinsic side jump plus their disorder renormalizations, and the last two terms represent the contributions due to skew scattering in the Born approximation and beyond it respectively. An important unprecedented finding of our study is that neither extrinsic spin-orbit coupling nor magnetic impurity scattering give a driving term in the kinetic equation \textit{parallel} to $H_{0{\bm k}}$ (that is, $\parallel {\bm \Omega}_{\bm k}$), thus the AHE response does not contain a term of order $[\hbar/(\varepsilon_F\tau)]^{-1}$ due to extrinsic spin-orbit scattering or magnetic impurity scattering. Such a term would have overwhelmed the intrinsic topological term in the ballistic limit. Since the intrinsic topological term is also of order $[\hbar/(\varepsilon_F\tau)]^{0}$, we conclude that there is no term $\propto n_i^{-1}$ (i.e. $\propto \tau$ only) in the AHE response of TI. 

In TI the spin and charge degrees of freedom are inherently coupled and the AHE current can also be viewed a steady-state in-plane \textit{spin} polarization in a direction \textit{parallel} to the electric field. The (bare) topological term has two equal contributions, which are part of the correction to the density matrix \textit{orthogonal} to the effective Zeeman field, therefore they represent an electric-field induced displacement of the spin in a direction transverse to its original direction. They are also obtained in the Heisenberg equation of motion if one takes into account the fact that ${\bm k}$ is changing adiabatically. Physically, the $\hat{\bm x}$-component of the effective Zeeman field ${\bm \Omega}_{\bm k}$ is changing adiabatically, and the out-of-plane spin component undergoes a small rotation about this new effective field. Consequently, each spin acquires a steady state component parallel to ${\bm E}$, which in turn causes ${\bm k}$ to acquire a small component in the direction perpendicular to ${\bm E}$. Elastic, pure momentum scattering [contained in $\hat{J}_0 (f_{\bm k})$] reduces this spin polarization because, in scattering from one point on the Fermi surface to another, the spin has to line up with a different effective field ${\bm \Omega}_{\bm k}$. The extra spin component of each electron is $\propto M$, however the final result is independent of $M$, as the integrand contains a monopole located at the origin in ${\bm k}$-space \cite{Culcer_AHE_PRB03, ZangNagaosa_TI_Monopole_PRB10}. As $M\rightarrow 0$ the effect disappears, since the correction to the orthogonal part of the density matrix vanishes. Finally, though the form of this term would hint that it is observable for infinitesimally small $M$, it was tacitly assumed that $M$ exceeds the disorder and thermal broadening.

Experimentally, for charged impurity scattering the figures depend on the Wigner-Seitz radius $r_s$, representing the ratio of the Coulomb interaction energy and the kinetic energy, with $\epsilon_r$ the relative permittivity. For Bi$_2$Se$_3$ with $r_s = 0.14$ \cite{Culcer_TI_Kineq_PRB10}, the dominant term by far is $\sigma_{yx}^{int} \approx - 0.53 \, (e^2/2h) \approx -e^2/4h$. We can consider also the (artificial) limit $r_s \rightarrow 0$, which implies $\varepsilon_r \rightarrow \infty$, artificially turning off the Coulomb interaction. As $r_s \rightarrow 0$, the prefactor of $-e^2/2h$ tends to $0.61$, while at $r_s = 4$, the limit of RPA in this case, it is $\approx 0.12$. Interestingly, for short-range scattering, the AHE current changes sign, with $\sigma_{yx}^{int} \approx 0.18 \, (e^2/2h)$. We note that $\lambda$ is not known for TI, yet that does not affect the central result of this work, since the intrinsic term is dominant. We have not taken into account the term in $\hat{J}_{ss}^{3rd}$ linear in ${\bm E}$, comparable in magnitude to the side jump.

We recall that the result presented above represents the contribution from the conduction band, and in principle an extra $e^2/2h$ needs to be added to the total result to obtain the signal expected in experiment. Since the conduction band is expected to contribute $\approx -e^2/4h$, this does not make a difference in absolute terms. Comparison with experiment must await further advances in materials growth, yet the most recent developments offer renewed grounds for hope. \cite{Kim_TI_Gate_MinCond_11}

\section{Summary}

We have determined the AHE conductivity due to the surface conduction band of 3D TI, including all intrinsic and extrinsic contributions. We have identified a skew scattering term in the Born approximation and one of third order in the scattering potential, as well as an intrinsic side-jump term and a side jump scattering term. The intrinsic topological term, renormalized by disorder, is dominant and of the order of $e^2/4h$, though non-universal, and easily observable experimentally. This finding provides an unmistakable signature of surface transport in TI.

DC acknowledges the partial support of the Chinese Academy of Sciences. We are grateful to N.~A.~Sinitsyn, Z.~Fang, J.~R.~Shi, W.~M.~Liu, X.~Dai, C.~G.~Zeng, Y.~Q~Li and A.-P. Li for enlightening discussions.

\end{document}